\documentstyle[epsfig,preprint,amssymb]{acmconf}

\newcommand{\eps}{\varepsilon}
\newcommand{\bh}{\overline{h}}

\newcommand{\wsat}{w_{\rm sat}}
\newcommand{\cP}{{\cal P}}

\begin{document}
\title{{\huge\bf\sf Optimal Detection of Sequence Similarity 
by Local Alignment}
}


\author{Terence Hwa\cr
Department of Physics\cr 
University of California at San Diego\cr
9500 Gilman Drive\cr
La Jolla, CA 92093-0319\cr\cr
E-mail: hwa@ucsd.edu
\and Michael L\"assig\cr
Max-Planck Institut f\"ur Kolloid-\\ und Grenzfl\"achenforschung,\cr
Kantstr. 55\cr 
14513 Teltow, Germany\cr\cr
E-mail: lassig@arktur.mpikg-teltow.mpg.de
}

\toappear{
\hrule
\vspace{\baselineskip}
To appear in {\it Proceedings of the Second Annual Int'l Conference
on Computational Molecular Biology}, 1998.\\
\vfill
Related (p)re-prints at
{\it http://matisse.ucsd.edu/$\sim$hwa}.
\vspace{\baselineskip}
}

\maketitle

\section*{ABSTRACT}

The statistical properties of local alignment algorithms with gaps
are analyzed theoretically for uncorrelated and
correlated random sequences. In the vicinity of the log-linear
phase transition, the statistics of alignment with gaps is shown to be
characteristically different from that of gapless alignment.
 The optimal scores obtained for
uncorrelated sequences  obey certain robust scaling laws.
Deviation from these scaling laws signals sequence homology, and 
can be used to guide the empirical selection of scoring parameters for
the optimal detection of sequence similarities. This 
can be accomplished in a computationally efficient way  
by using a novel approach focusing on the score profiles. 
Furthermore, by assuming a few gross features characterizing
the statistics of underlying sequence-sequence correlations,
quantitative criteria are obtained for the 
choice of optimal scoring parameters: Optimal similarity detection
is  most likely to occur in a region close to the log side of 
the log-linear phase transition.

%
%


\vspace{\baselineskip}
\noindent {\ninsfb Keywords:} \, 
sequence alignment;
homology; optimization;
phase transition





\section{INTRODUCTION}

Sequence alignment is a vital tool in molecular biology. It has been used
extensively in discovering and understanding the functional and evolutionary
relationships among genes and proteins~\cite{review,msw2}. There are two basic
types of alignment algorithms: algorithms without gaps, e.g.,  BLAST and\\
FASTA~\cite{gapless}, and algorithms with gaps, e.g., the
Needleman-Wunsch algorithm~\cite{nw} and the Smith-Waterman algorithm~\cite
{sw}. Gapless alignment is widely used in large-scale database searches 
because the algorithms are fast~\cite{gapless}, the results depend only 
weakly on the choice of
scoring systems~\cite{altschul}, and the statistical significance of the
results is well-characterized~\cite{amw,karlin1,karlin2}. However, gapless
alignment is not sufficient for the detection of {\em weak} sequence
similarities~\cite{pearson}. For the detailed analysis of such sequences, 
algorithms with gaps are necessary~\cite{gotoh,msw1,msw2}.
Advancing our understanding of the statistics of
gapped alignment could therefore be critical to the wider usage of these
more powerful tools.

A notorious difficulty for any alignment is the
selection of scoring schemes and/or parameters: In a generic sequence
matching problem, a score is assigned to each alignment of given sequences,
based on the total number of matches, mismatches, gaps, etc. Maximization of
this score defines an optimal alignment. However, it is well known that the
optimal alignment of given sequences strongly depends on the particular
scoring scheme and/or parameters used. Consequently, the {\em fidelity} of
an alignment, i.e., the extent to which the alignment captures mutual
correlations among the aligned sequences, can depend strongly on the choice
of scoring parameters. Understanding the influence of these parameters on
the resulting alignment and choosing the appropriate parameters are
therefore important for the proper application of these algorithms. This
 requires the knowledge of the statistics of alignment results,
which has been obtained only for  gapless 
alignments~\cite{altschul,amw,karlin1,karlin2}. 
For alignments with gaps, appropriate parameters have so far been
chosen mostly by trial and error, although there have been systematic
efforts to establish a more solid empirical footing~\cite{benner,vw}.

Recently~\cite{hl,dhl} we have  
analyzed the statistical properties of {\em global} alignment
with gaps. Such algorithms align sequences of similar
 lengths in their entirety. By exploiting mathematical analogies to 
certain well-studied problems of statistical mechanics~%
\cite{kardar,hf,hn,kl}, 
we have obtained a  quantitative description of the global alignment
statistics  for mutually {\em %
uncorrelated} as well as {\em correlated} sequence pairs. 
Here we  extend the analysis to  {\em local}
alignment algorithms~\cite{sw} which 
find the best match between {\em contiguous subsequences}, subject to 
(finite) penalties for gaps and mismatches. 
For uncorrelated random sequences, i.e., for independent sequences
with iid or Markov letters, it is well known that
depending on the choice of scoring parameters, the length of the optimal
subsequence alignment depends either linearly or logarithmically on the
total length of the sequences~\cite{wga,aw}. A phase transition line separates 
the space of
scoring parameters into two regimes: the ``linear phase'' for small gap and
mismatch costs, and the ``log phase'' for large penalty costs. It is clear
that local alignment deep in the linear phase is equivalent to global
alignment (and hence described by the results of our previous studies). 
On the other hand, the log phase at high gap penalty becomes indistinguishable
from the log phase of gapless alignment. Indeed there have been
extensive  empirical efforts~\cite{swb,ccl,mott,wv2,wv3,altschul2}
to characterize the statistics of the log phase of gapped algorithms
by an effective description as gapless alignment with modified parameters.
While this approach is reasonable far away from the phase transition,
it becomes questionable as the phase transition
line is approached, since the linear phase itself is completely different
from the log phase. On the other hand, 
the loci of scoring parameters for optimal similarity detection
appear to lie in the log phase close to the phase transition line,
 according to recent empirical studies by Vingron and 
Waterman~\cite{vw}. Hence, understanding the log-linear
phase transition is crucial for optimizing the  detection of 
sequence similarity and quantifying its statistical significance.

In this work, we apply the well-established theory of phase transition~%
\cite{phase-trans} to the log-linear transition of gapped local alignment. 
We find various statistical properties at and in the vicinity of the 
transition to be governed by {\em scaling laws} analogous to those
recently discovered for global alignment~\cite{hl,dhl}. The transition turns 
out to differ qualitatively and quantitatively from its counterpart 
in gapless alignments. 
Our results lead to quantitative criteria for the optimal choice
of scoring parameters, given certain gross statistical characteristics 
of the expected sequence correlation. 
In particular, they explain why optimal parameters
for weakly correlated sequence pairs are in the vicinity of the
phase transition line as observed by Vingron and Waterman~\cite{vw}. 
Also emerging from this work is a versatile method to
detect sequence correlation and characterize its statistical significance
empirically for  sequences with {\it a priori} unknown correlations.



\section{REVIEW OF ALIGNMENT ALGORITHMS}

We study the Smith-Waterman (SW) local alignment algorithm applied to a pair
of long nucleotide sequences ${\cal P}_1$ and ${\cal P}_2$, with lengths $%
N_1\simeq N_2\gg 1$. Let $P_{n,i}\in \{A,T,G,C\}$ be the $i^{{\rm th}}$
element of the sequence ${\cal P}_n$. A particular alignment consists of an
ordered set of pairings of two elements $(P_{1,i},P_{2,j})$, or of an
element with a gap, for any contiguous subsequence of length $\ell _1\leq N_1$
in sequence ${\cal P}_1$ and length $\ell _2\leq N_2$ in sequence ${\cal P}_2$
[see Fig.~1]. The simplest scoring system assigns a positive score of $+1$ 
if the two elements paired together are identical, and a negative score of 
$-\mu $ if the two are different. Each pairing of an element with a gap is
penalized with a negative score $-\delta $. (In the simple case considered
here, we shall not distinguish between gap initiation and gap extension.
It is then sufficient to consider only  the region 
$2\delta \ge \mu$\footnote{For $2\delta < \mu$, it is 
always favorable to replace a mismatch by two gaps, so the outcome
of an alignment becomes independent of $\mu$.}.)
The sum of the scores of all individual pairings of a given alignment is the
total score for that alignment. An {\em optimal} alignment is one for which
the total score is maximized for a given set of scoring parameters 
$(\mu,\delta )$. The SW algorithm uses the dynamic programming method 
to find the optimal alignment of all possible subsequences.
Key to the algorithm is the unique representation of an alignment by a
directed path (see Ref.~\cite{nw} and Fig.~1). Let us briefly recall the
algorithm below, cast in a slightly different notation to facilitate the
subsequent analysis.

\begin{figure}
\centerline{\epsfig{file=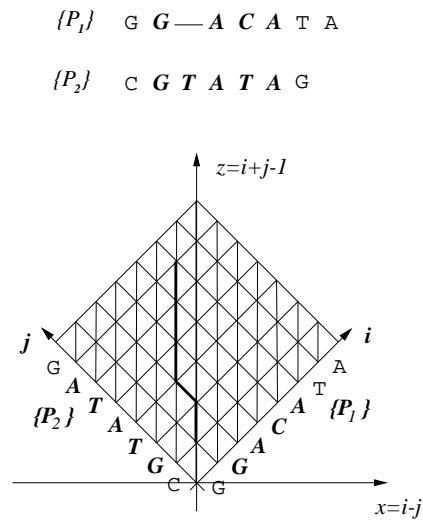, width=2.2in, angle=0}}
\vspace{\baselineskip}
\caption{
One possible local alignment of the two nucleotide sequences, $\cP_1=
GGACATA...$ and $\cP_2=CGTATAG...$ The aligned subsequences are shown
in boldface, with $4$ pairings (three matches, one mismatch)  and one gap.
This alignment can be represented uniquely as a directed path on
the alignment lattice; each left (right) turn of the path correspond
to a gap insertion in sequence $\cP_1$ ($\cP_2$).
}
\end{figure}

Consider the alignment lattice shown in Fig.~1. We label each lattice point by
the coordinate $(x,z)$, with the lower tip of the lattice anchored
at $(0,0)$. The 
highest total score of all alignment paths  ending at a point $(x,z)$ 
is denoted by $h(x,z)$. Given $h(x,z)$ and $h(x,z-1)$ for all $x$, 
$h(x,z+1)$ can be computed from the recursion relation 
\begin{equation}
h(x,z+1)=\max \left\{
\begin{tabular}{l}
$h(x+1,z)-\delta$ \\
$h(x-1,z)-\delta$\\
$h(x,z-1)+u(x,z)$\\
$h_0$
\end{tabular}
\right\},
\label{rule}
\end{equation}
with $\delta $ being the gap insertion cost, $u$ being the match/mismatch
score to be specified below, and $h_0$ being a cutoff score. The SW
algorithm has $h_0=0$, which effectively deletes the
segment of the alignment path connecting $(0,0)$ and $(x,z)$ if $h(x,z)\le 0$.
In contrast, the global alignment algorithm of Needleman and Wunsch has no
cutoff, corresponding to the limit $h_0\to -\infty $. Also,
gapless local alignment (with $\delta \to \infty$) corresponds to the limit
of the recursion relation (\ref{rule}) involving only one value of $x$,
say  $x=0$. The scoring function 
$u(x,z)$ gives the match/mismatch score of aligning the elements $P_{1,i}$
with $P_{2,j}$, with $x=i-j$ and $z=i+j-1$. For simplicity, we use in this
study the form
\begin{equation}
u(x=i-j,\,z=i+j-1)=\left\{ 
\begin{tabular}{ll}
$1$ \qquad  & if $\quad P_{1,i}=P_{2,j}$ \\ 
$-\mu $ \qquad  & if $\quad P_{1,i}\ne P_{2,j}$%
\end{tabular}
\right. .  \label{score}
\end{equation}
More elaborate forms of the scoring function are easily incorporated and do
not change key results of this study.

If $z$ and $x$ are regarded as ``time'' and ``space'' variables, respectively,
the recursion relation (\ref{rule}) can be viewed as a
``dynamical process'' describing the time evolution of the
one-dimensional ``score profile'' $h(x,z)$.  
(Similarly, gapless local alignment involving only the site $x=0$
corresponds to the ``zero-dimensional'' limit.)
This dynamic analogy will be pivotal in guiding the ensuing analysis. 


\section{GLOBAL ALIGNMENT}

\subsection{Statistics of Uncorrelated Sequences: Universal Scaling Laws}

Let us first review some of the relevant results we previously obtained for the
{\em global} alignment of random sequence pairs~\cite{hl}. 
In Fig.~2(a), we show several
representative constant-$z$ slices of the score profile $h(x)$ obtained by
iterating Eq.~(\ref{rule}) with $(\mu ,\delta )=(0.5,2.0)$.
 The alignment
algorithm is applied to one pair of random sequences each of length $N=10000$.
Results are shown for a central rectangular 
region\footnote{Due to boundary effects arising from the diamond-shaped
alignment lattice (Fig.~1), the total score $h(x,t)$ certainly decreases
(quadratically on average) as one moves far away from the center at $x=0$. 
To remove these spurious effects, we  focus
our attention only to the central strip of the alignment lattice,
e.g., for $-X/2\le x\le X/2$ and $X/2\le z \le 2N-X/2$, with $X\lesssim N$.
All results reported here are obtained using this
strip geometry. For long sequences, the statistics of the score profile
obtained from the strip is indistinguishable from that obtained with 
the full alignment lattice.} 
of the alignment lattice, 
$-X/2\le x\le X/2$ and $X/2\le z\le 2N-X/2$ with $X \lesssim N$, 
starting from the initial condition 
$h(x,z=X/2)=0$. 
It will be convenient to use a shifted time-like variable, $t\equiv z-X/2$. 
In Fig.~2(a), we see a series of disorderly score profiles, with the 
``spatial'' average 
\begin{equation}
h(t)=X^{-1}\sum_{x=-X/2}^{X/2}h(x,t)
\end{equation}
advancing
steadily in $t$. For large $t$, we  
obtain the linear dependence, $h (t)=v_0(\mu ,\delta )t$ (not shown).
The value of the rate $v_0$ itself is not important for global alignment.
(Thus, $h(t)$ could as well be {\em decreasing} linearly in $t$.)
More significant is the spatial variation in the profile which always
increases for increasing $t$. This is more clearly seen by plotting the
effective width of the profile, $w(t)$,  defined as
\begin{equation}
w^2(t)=\frac 1X\sum_{x=-X/2}^{X/2}\left[ h(x,t)- h(t)\right] ^2,
\label{width}
\end{equation}
or alternatively, the difference between the highest and lowest point of the
profile, $\Delta h(t)=h_{{\rm max}}(t)-h_{{\rm min}}(t)$; see Fig.~2(b).
\begin{figure}[t]
\centerline{\quad\epsfig{file=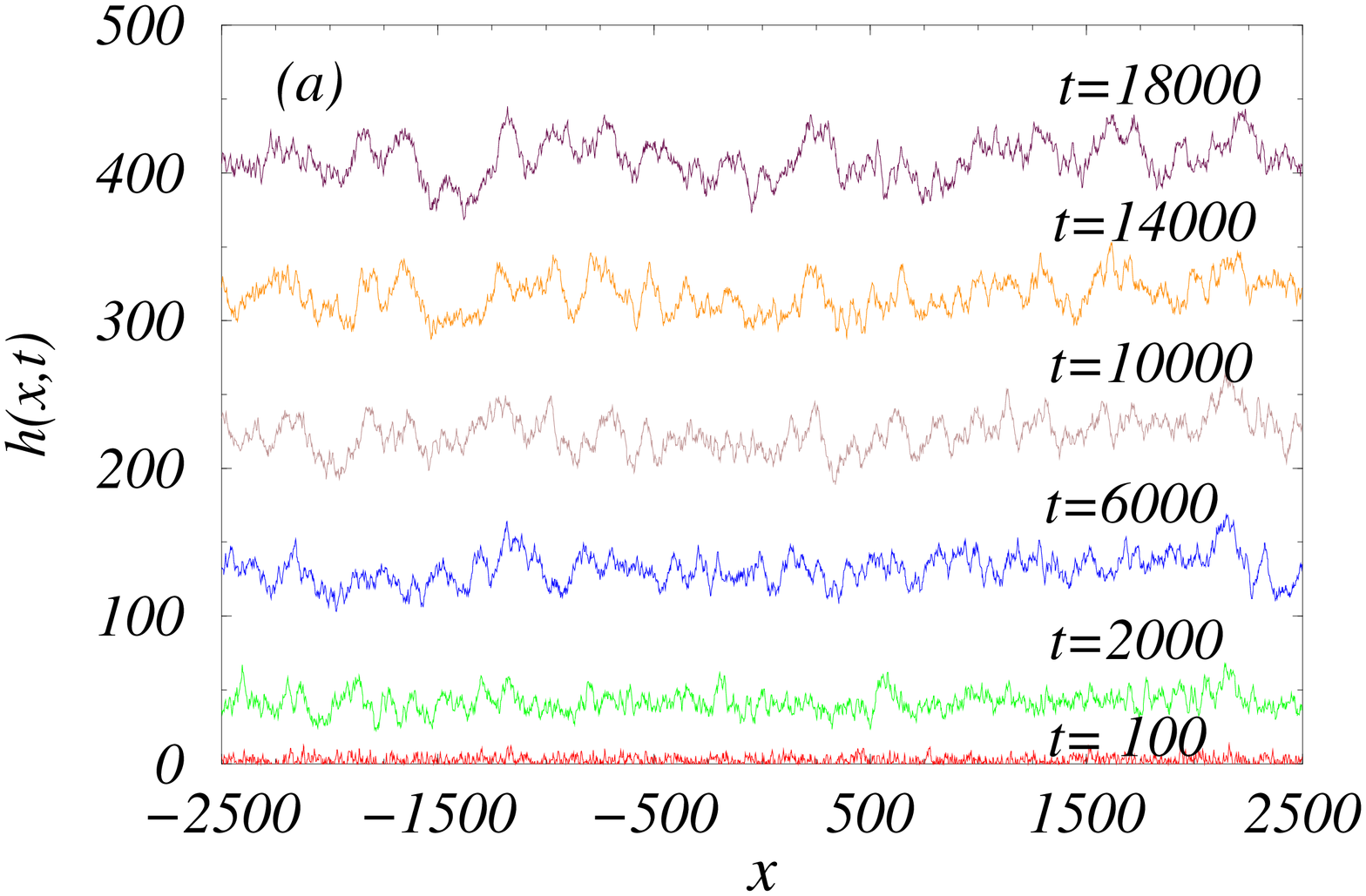, height=3.0in, angle=270}}
\centerline{\epsfig{file=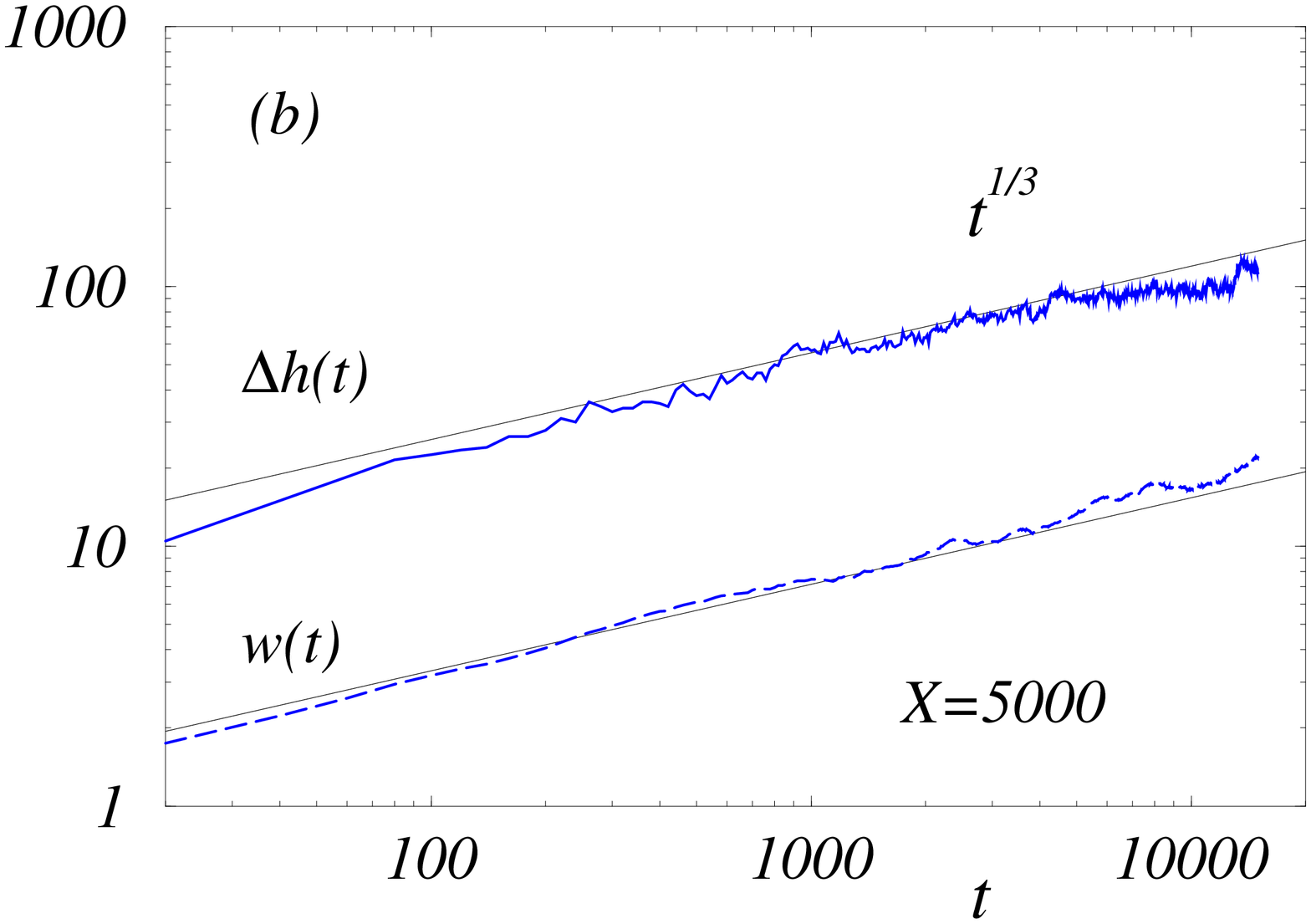, height=2.8in, angle=270}}
\caption{
(a) A typical series of score profiles $h(x,t)$ obtained
from the global alignment of a pair of uncorrelated random sequences;
parameters used are $(\mu,\delta)=(0.5,2.0)$. 
(b) Gradual increase in the ``roughness'' of the profile, 
as characterized by either the width
$w(t)$ or the range $\Delta h(t)$ over the range of $x$ shown in (a). 
The straight lines indicate the suggested power law dependence.
}
\end{figure}

The roughness of the profile, as quantified by either $w(t)$ or $\Delta h(t)$,
is an important characteristic of the alignment, since it indicates how
strongly the optimal alignment dominates over  the suboptimal alignments.
The  score profiles of Fig.~2(a) show  the {\em %
weak dominance} of the optimal alignment and the existence of a large number
of suboptimal alignments. The statistics of these suboptimal alignments
has been recognized recently as a valuable tool in sequence 
alignment; for an interesting recent exposition, see Ref.~\cite{vingron}. 
~From Fig.~2(b), it appears that the roughness grows with a sub-linear power
in $t$. This is verified in Fig.~3(a), 
where we show the ensemble average $\overline{w}(t)$ 
for different sets of scoring parameters $(\mu ,\delta )$, each curve
averaged over $1000$ pairs of uncorrelated random sequences. (Throughout the
text, we use overbars to denote averages of an ensemble of
random sequence pairs.) It is seen that for large $t$, the width obeys the
asymptotic scaling law 
\begin{equation}
\overline{w}(t) = B(\mu ,\delta )\,t^\omega .
\label{width.scale}
\end{equation}
Different parameter choices only affect the prefactor $B$, but not the
exponent $\omega \approx 1/3$. The same scaling law (with a larger
coefficient) is found also for $\overline{ \Delta h(t)} $.
\begin{figure}[h]
\centerline{\epsfig{file=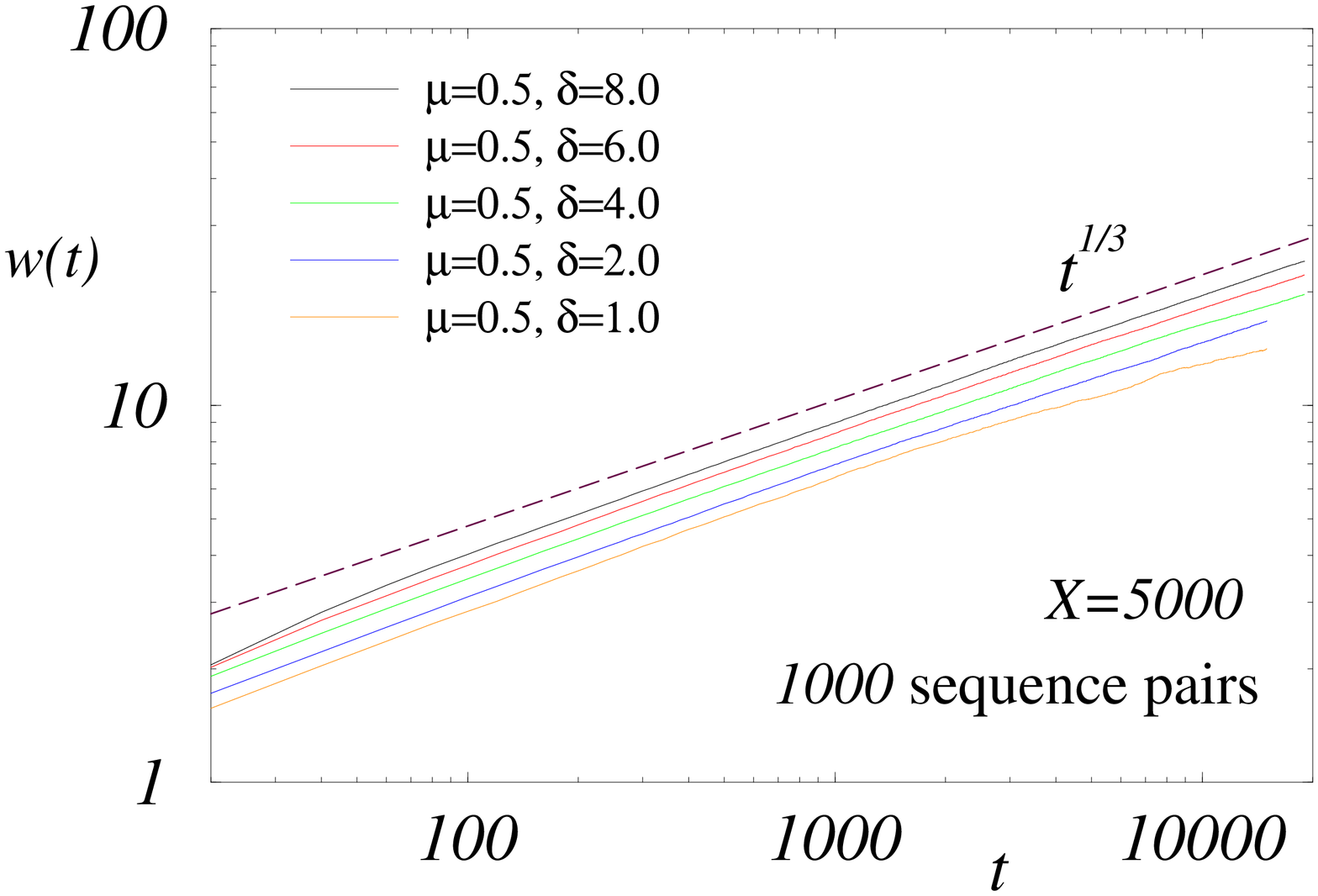, height=3.0in, angle=270}}
\centerline{\epsfig{file=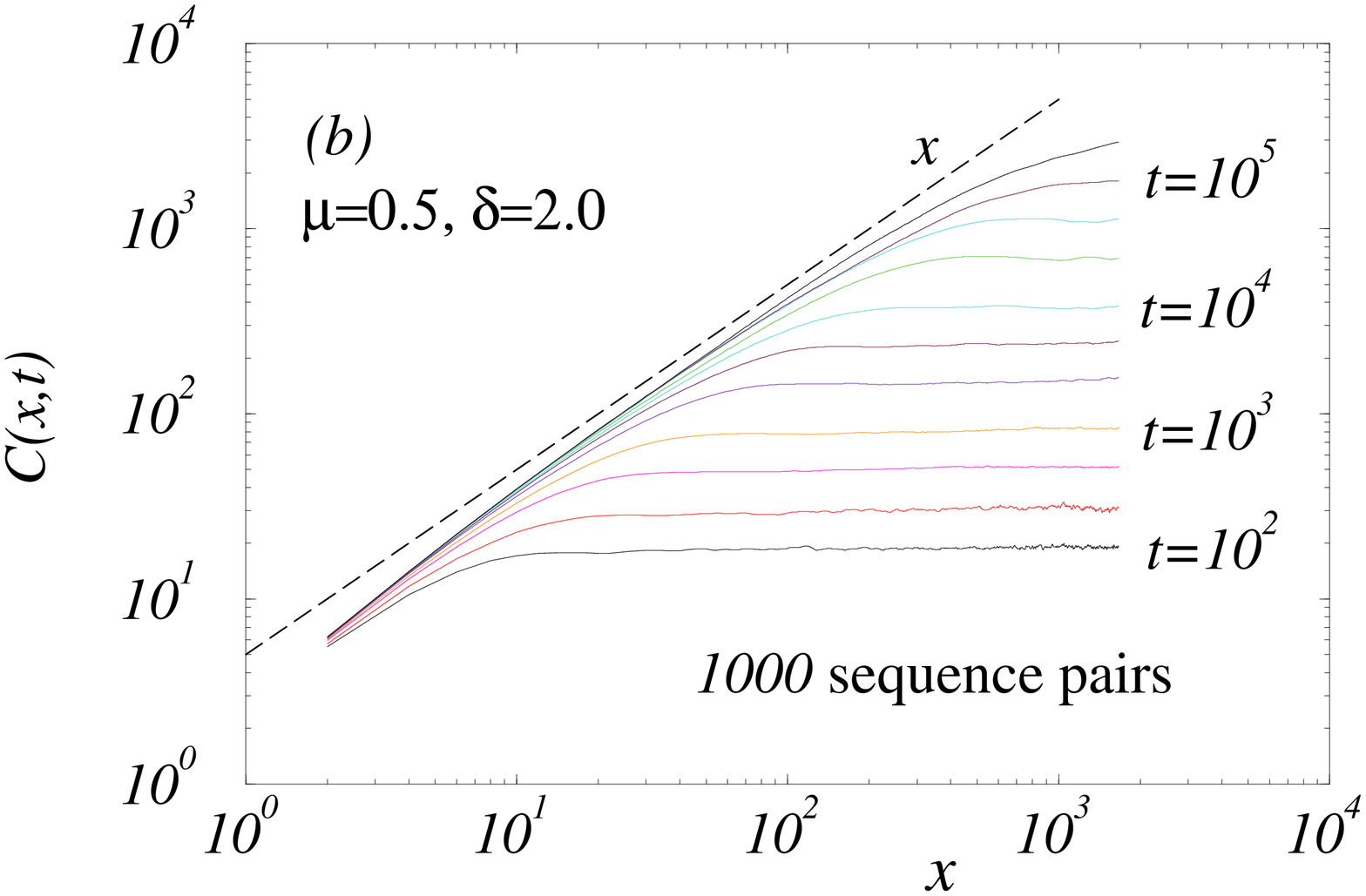, height=3.0in, angle=270}}
\caption{
(a) The ensemble averaged roughness (width) for different sets of scoring 
parameters, each  averaged over $1000$ pairs of
randomly generated sequences of length $N=10,000$.
The dashed straight line indicates the expected power law form.
(b) The ensemble averaged spatial correlation function $C(x,t)$ 
for $(\mu,\delta)=(0.5,2.0)$. The dashed line indicates the expected scaling
behavior at large $t$. For small $t$'s, $C(x,t)$ saturates to the order
of $\overline{w}^2(t)$.
}
\end{figure}

The scaling law (\ref{width.scale}) is an example of the ``universal scaling
phenomena'' well studied in statistical mechanics~\cite{phase-trans}. 
The exponent $\omega $ is
a very robust characteristics of random sequence alignment. 
It quantifies the
roughness of the profile and hence the dominance of the optimal alignment.
It does not depend on details of the scoring function, (e.g., 
whether or not gap initiation and extension are differentiated) but
only on a few qualitative characteristics such as the number of 
sequences aligned or the type of the correlations between them.

In Ref.~\cite{hl}, we have given arguments suggesting that the result 
$\omega =1/3$ is exact for the global alignment of random sequence pairs. Our
approach is based on the close analogy of the recursion relation (\ref{rule})
and discrete models of kinetic surface growth studied extensively in
statistical mechanics in the past decade\footnote{General reviews of the
kinetic growth problem and its relation to the problem of first passage
percolation can be found in Refs.~\cite{ks} and \cite{hhz}.}.
In these growth models, $h(x,t)$ describes the height
profile of a one-dimensional surface at time $t$. The growth is driven
by a stochastic process that governs  the deposition and removal of
material on the surface and is described by the random variable
 $\eta (x,t)= \frac 12u(x,t)-\delta $. 
  The large scale behavior of the profile $%
h(x,t)$ generated by the growth model is well described by the {\em %
differential} growth equation~\cite{kpz}
\begin{equation}
\frac{\partial h}{\partial t}=\nu \frac{\partial ^2h}{\partial x^2}+\frac 
\lambda 2\left( \frac{\partial h}{\partial x}\right) ^2+\eta (x,t),
\label{kpz}
\end{equation}
with the coefficients $\nu $ and $\lambda $ being functions of $\delta $ and 
$\mu $. Eq.~(\ref{kpz}) is known as the 
Kardar-Parisi-Zhang equation, closely related to the
noise-driven Burgers' equation~\cite{fns}. 
If $\eta (x,t)$  is an uncorrelated
Gaussian random variable,  the stationary state ($t \to \infty$)
of the surface can be obtained  exactly~\cite{ks}, resulting
in the Gaussian equal-time distribution 
\begin{equation}
P\left\{ h(x,t\rightarrow \infty )\right\} \propto e^{-\frac 1{2D}%
\sum_x\left[ h(x+1,t)-h(x,t)\right] ^2} ,  \label{P.h}
\end{equation}
and a coefficient $D(\mu ,\delta )$. Approaching the steady state, one has 
$\overline{w}(t\gg 1)=B\,t^\omega $, with $B\approx D^{2/3}$ and $\omega =1/3$ exactly.
For sequence alignment, the variables $\eta(x,t)$ and
$\eta(x', t')$ at different points are not independent given their definition
above. One can verify for
example that the higher moments of $\eta $ are long-range correlated. 
However, as argued in Ref.~\cite{hl} and [Hwa and L\"assig (to be published)], 
these correlations do not affect the asymptotic scaling
behavior\footnote{More detailed discussions are given in Ref.~\cite{ch}
in the context of a number of closely related physics problem.}.
 The close correspondence of the numerical
results on $w(t)$ (Fig.~3(a)) and the exact result $\omega =1/3$ supports this
conclusion. An independent check is to measure directly the equal-time
correlation of $h$, $C(x,t)= \overline{ [h(x+x^{\prime },t)-h(x^{\prime
},t)]^2} $. From Eq.~(\ref{P.h}), we expect to have $%
C(x,t)=D\,|x|$ for sufficiently large $t$ (or sufficiently small $x$). For
finite $t$, $C(x,t)$ must eventually saturate, to a value 
$\sim \overline{w}^2(t)$ for $x>O\left( t^{2/3}\right) $. 
The numerically obtained forms of $C(x,t)$ are
shown in Fig.~3(b) for $(\mu ,\delta )=(0.5,2.0)$. The results are in good
agreement with the anticipated form. Similar results are obtained for other
values of the scoring parameters.


\subsection{Correlated Sequences: Similarity Detection from the Score 
Profile}

To illustrate how the above knowledge can be used for the purpose of
similarity detection, we next describe the score profile for the global
alignment of {\em correlated} sequences. Sequence correlations are obtained
by first generating a random ``ancestor sequence'', and then making {\em %
imperfect} replicas  ${\cal P}_1$ and ${\cal P}_2$. The degree of
sequence-sequence correlation is quantified by the average number of point
substitutions per base $p_s$ and the average number of indels per base $p_t$
made in the replication of each daughter sequence. 
(See Refs.~\cite{hl,dhl} for details.) In Fig.~4(a), we show the score 
profile $h(x,t)$ and the range $\Delta h(t)$ for a pair of heavily mutated 
daughter sequences with $p_s=40\%$
and $p_t=20\%$, corresponding to a fraction of  $<30\%$ conserved elements.
The score profiles obtained are very different from
the profiles characteristic of uncorrelated random sequences shown in 
Fig.~2(a). For
correlated sequences (Fig.~4(a)), a  peak emerges from the disordered
background after some time. The height of the
peak advances steadily at a rate $v$ which exceeds the growth rate of the
background $v_0$. Thus, the peak gradually broadens to
engulf the entire profile. The dominance of the central peak reflects the
existence of {\em strongly preferred} alignments for correlated sequences,
in marked contrast to the alignment of random sequences\footnote{In statistical
mechanics, one considers the free energy, which is the negative of the score.
The score profile discussed here is a measure of the ``energy landscape''.
A peak in the score profile corresponds to a valley or a ``funnel'' in
the energy landscape. It has been suggested that the energy landscape
of heteropolymeric systems such as a protein has the funnel 
shape~\cite{funnel}. The score profile (Fig.~4(a)) obtained here is
the first known example  of a large heteropolymeric system for which
the suggested landscape is directly observed.}.

The existence of
the peak can be taken as a manifestation of inter-sequence correlations. The
strength of sequence correlation detected is quantified by the
difference between the growth rate of the peak and the background, 
$\varepsilon \equiv v-v_0$. The magnitude of $\varepsilon $ is clearly
dependent on the mutation rates $p_s$ and $p_t$, but also depends on the
scoring parameters $\mu $ and $\delta $. The functional form of 
$\varepsilon(\mu,\delta;p_s,p_t)$ has recently been investigated 
in detail~\cite{dhl} and will not be addressed here.

 A peak in the score profile is discernible from
the background  only if the height of the peak, of the order 
$\varepsilon \cdot t$ after $t$ steps, exceeds the roughness of the
background $\sim B\cdot t^{1/3}$(see Fig.~4(b)). Hence, using the
score profile approach, one can detect 
correlations between sequences if their lengths exceed the threshold length 
\begin{equation} 
t_c  (\mu ,\delta ;p_s,p_t)\sim \left[ B(\mu ,\delta )/\varepsilon (\mu
,\delta ;p_s,p_t)\right] ^{3/2} \label{cond2}.
\end{equation}
Minimization of $t_c$ (for a given sequence pair) is a natural
empirical criterion for optimal similarity detection, 
yielding a preferred choice of scoring parameters
 $\delta^*(\mu;p_s,p_t)$.
A more fundamental criterion for choosing the optimal scoring parameters 
is to maximize the {\em fidelity} of the alignment, i.e.,  
the extent to which the optimal alignment reconstructs the 
ancestor sequence~\cite{hl}. The dependence
of fidelity on scoring parameters has been studied in detail in 
Ref.~\cite{dhl}. It was found that maximizing the
fidelity is closely related (and equivalent for practical purposes)
to minimizing the threshold length $t_c$. Thus, the empirical criterion
based on the score profile is indeed a versatile way of locating the
optimal parameters.

\begin{figure}[t]
\centerline{\quad\epsfig{file=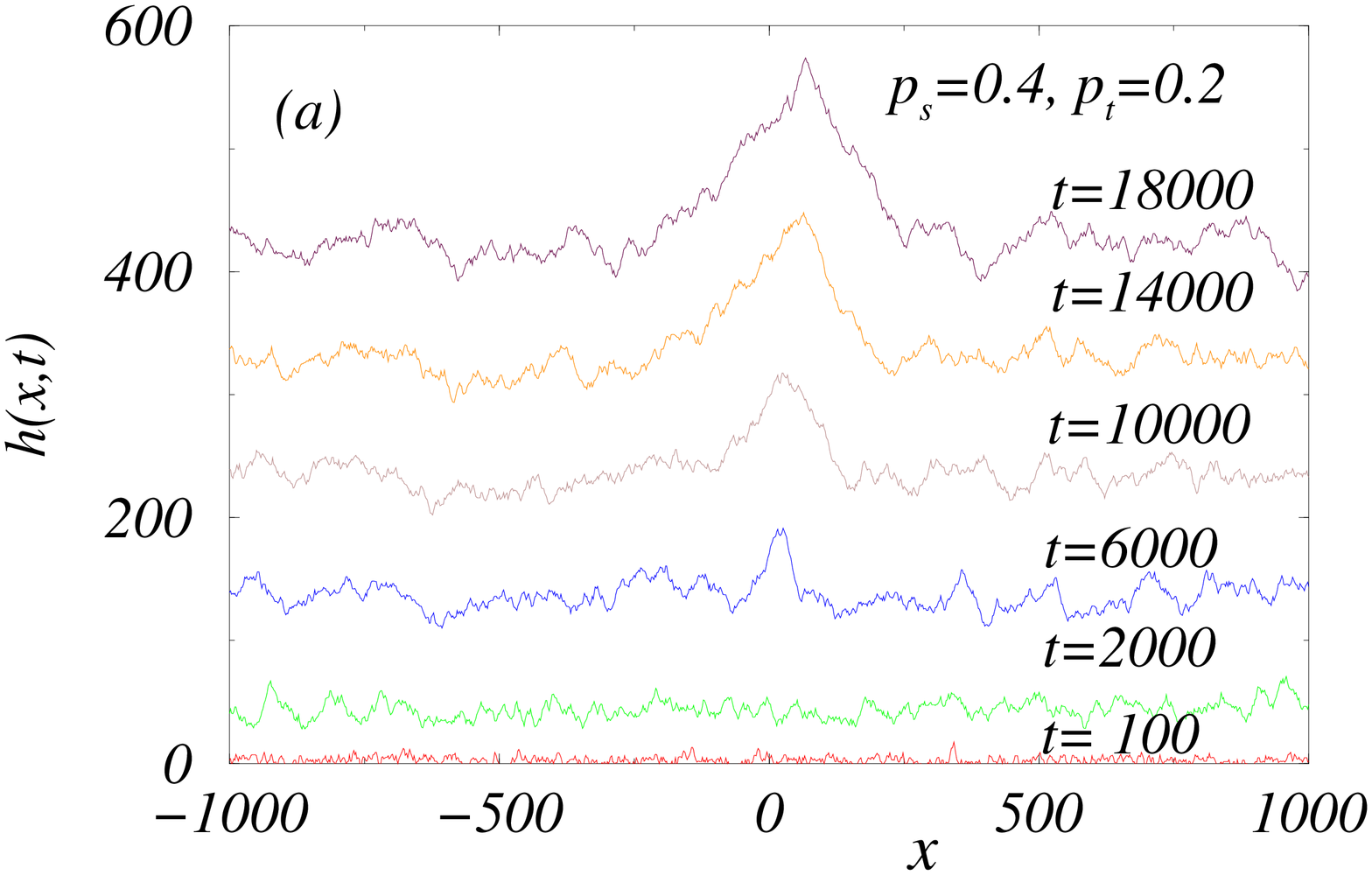, height=3.0in, angle=270}}
\centerline{\epsfig{file=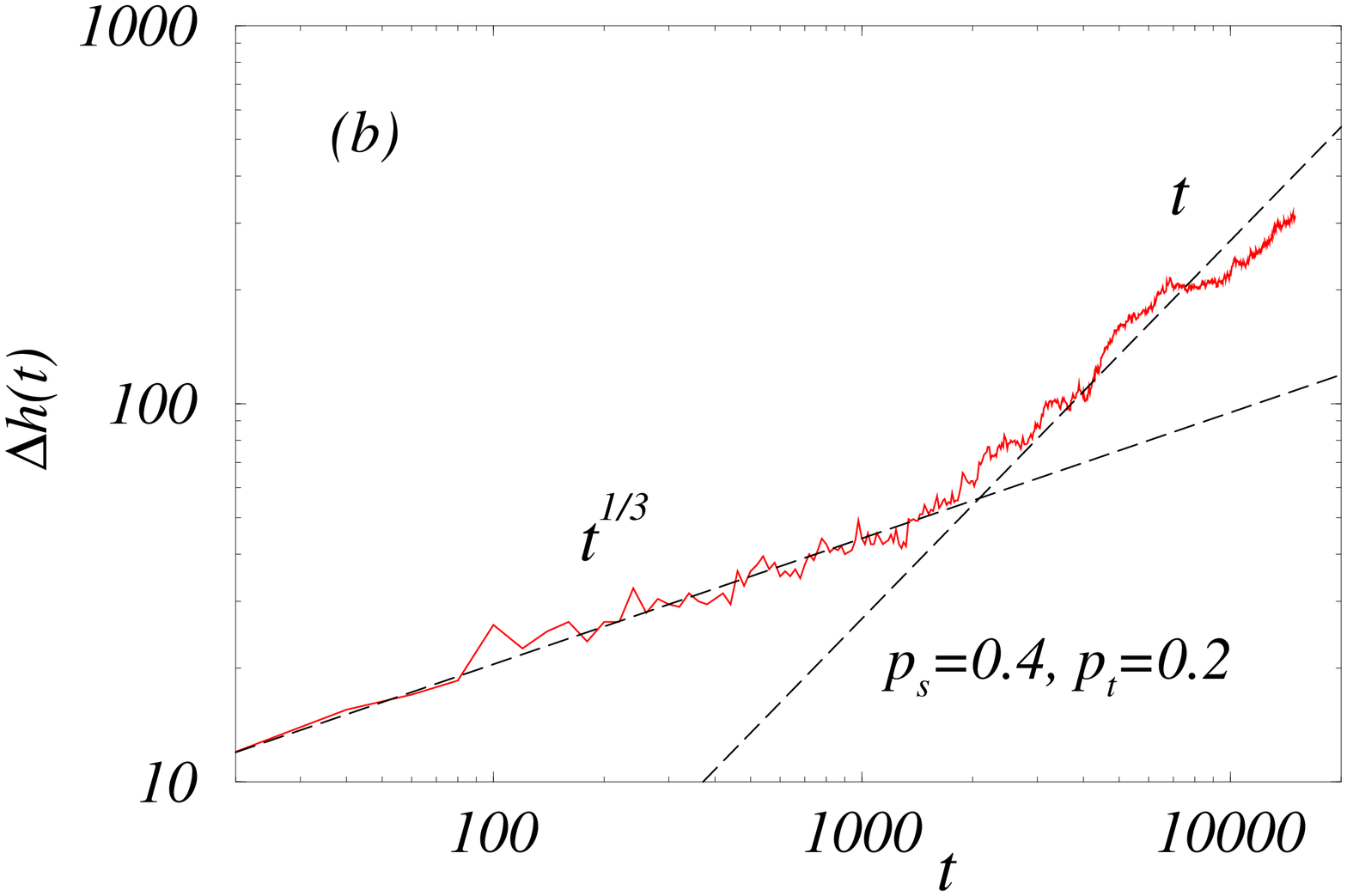, height=3.0in, angle=270}}
\caption{
(a) The score profiles
obtained from the global alignment of a pair of
weakly correlated sequences (with conserved fraction $<30\%$)
are characteristically different in appearance 
from those of the uncorrelated sequences shown in Fig.~2(a).
(b) The roughness corresponding to the profiles of (a)
begins to deviate from the $t^{1/3}$ behavior  at $t \simeq 2000$.
}
\end{figure}

The above strategy of similarity detection 
is close in spirit to the well-known ``shuffling method'', 
which  compares the alignment score of a given sequence pair to the
distribution of scores obtained from  aligning the ensemble of randomly
shuffled sequences (for the same set of scoring parameters). 
However, by making use of the spatially-extended 
score profile and its time evolution (as opposed to keeping track of only
the optimal score), we have demonstrated that this comparison 
can be accomplished by one {\em single} alignment. 
This is a drastic improvement over 
the shuffling method, which requires  the 
generation and alignment of an {\em ensemble} of sequences.
It should thus be possible to minimize  $t_c(\mu,\delta) $ 
empirically for sequence pairs with {\it a priori} unknown correlation,
since the value of $t_c$ can be obtained directly from the onset of
score peak shown in Fig.~4(b), without any assumption on the nature
of sequence correlations. It would be particularly interesting
to combine this approach  with the efficient ensemble and parametric
alignment algorithms~\cite{wel,gusfield,msw94} 
which find optimal scores for all parameters.


\section{LOCAL ALIGNMENT WITH GAPS}

We now describe the statistical properties of local alignment, with $h_0=0$
in the recursion relation (\ref{rule}). Local alignment is necessary since
 often  only a subsequence  of one sequence 
is correlated with a subsequence of another. 
Let the lengths of the correlated subsequences
be $\ell _1\approx \ell _2\approx \ell $.
If the positions of the subsequences were known, the correlations 
would be detectable by global alignment of these subsequences
if $\epsilon \ell > B \ell^{1/3}$, i.e.,
if $\ell > t_c$. However, if global alignment is applied to the 
entire sequences, the background noise $BN^{1/3}$ will in general 
overwhelm the score signal of the correlations $\epsilon \ell$.
The advantage of 
local alignment  is that by cutting off the length of the aligned segment,
it limits the background roughness to a 
{\em finite} value such that the correlation peak can
still be detected.


\subsection{Uncorrelated Sequences: the Log-Linear Phase Transition}

We first discuss the background, i.e., the local alignment of uncorrelated
random sequence pairs. Unlike global alignment, the outcome of local
alignment depends critically on the value of $v_0$, i.e., the growth rate
of the average score of the corresponding global alignment problem. 
For scoring parameters $(\mu ,\delta )$ in the regime where
 $v_0(\mu ,\delta ) > 0$, the score $h(x,t)$ grows without bound
 also for local alignment, and the existence of a
cutoff score $h_0=0$ is immaterial in the limit of large $t$. 
Thus the asymptotic properties of local and global
alignments are identical in this regime. (It acquires
the name ``linear phase'' since the average score $\overline{h}(t)$ advances
linearly by definition.) For $v_0 (\mu, \delta) < 0$, on the other hand,
the average $\overline{h}(t)$ saturates quickly to a
 constant value, and  the
largest value $h(x,t^{\prime }\leq t)$ scales as $\log (t)$ due to 
exponentially rare events. 
As has been pointed out by Arratia and Waterman [unpublished], 
the condition\footnote{However, finite-size effects~\cite{phase-trans}
may cause a slight shift in the apparent transition point. 
For example, the fits of 
 Figs.~6 and 7  found  $\mu_c \approx 0.7085$ for $\delta=2.0$,
whereas $v_0=0$ occurred at $\mu \approx 0.7040$ for $\delta=2.0$.}
\begin{equation}
v_0(\mu, \delta) = 0
\end{equation}
defines the phase transition line separating  the two regimes (see Fig.~5).
\begin{figure}
\centerline{\quad\epsfig{file=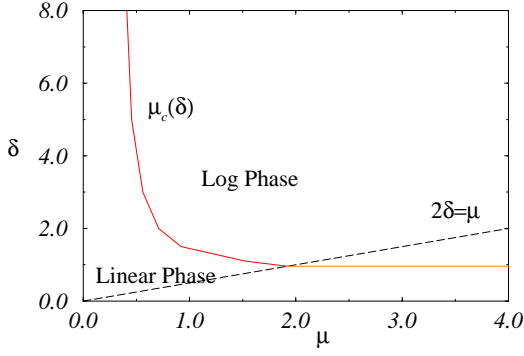, height=2.7in, angle=270}}
\caption{
The critical line $\mu_c(\delta)$ separating the log and the linear phases
is obtained from the condition $v_0(\mu_c,\delta)=0$. (The critical line
becomes independent of $\mu$ below the line $2\delta = \mu$.)
}
\end{figure}
The statistical properties in the vicinity of 
this phase transition have recently been studied in
the context of some related physics problems~\cite{mh}. We will briefly
summarize the results below.

The phase transition is very similar to that of gapless local alignment,
except that the $t^{1/2}$ score dependence at the transition
 is replaced by 
\begin{equation}
h_c(t)\equiv  \overline{h}(t)_{\mu =\mu _c}=b(\delta )\,t^{1/3},
\label{h.c}
\end{equation}
along the critical line $\mu _c(\delta )$
for large $t$, with a coefficient $b(\delta )\sim \,B(\mu _c(\delta ),\delta)$
dependent on the location along the phase transition line. 
Slightly away from the
critical line on the log side, i.e.,  for $\widetilde{\mu}\equiv
\mu -\mu _c\gtrsim 0$ where $v_0\propto -\widetilde{\mu}\lesssim 0$, $ 
\overline{h}(t) $ is indistinguishable from $h_c(t)$ for small $t$,
saturating eventually to a constant value 
$h_{\rm sat}\approx b^{3/2}/|v_0|^{1/2}$.
The saturation scale $t_\times$ is obtained from
the condition $h_c(t=t_\times) \sim h_{\rm sat}$, yielding 
\begin{equation}
t_\times \sim ( b/|v_0|) ^{3/2}\propto |\widetilde{\mu}|^{-3/2}.
\label{length}
\end{equation}
This is the length of the the optimally aligned subsequences 
selected by local alignment. For sequence lengths $N$ much longer than 
$t_\times$, it is reasonable to approximate the score statistics by that of 
{\em gapless} local alignment, as has been attempted 
previously~\cite{swb,ccl,mott,wv2,wv3,altschul2}. However, because 
$t_\times$ diverges  as the critical point ($\widetilde{\mu}=0$) is approached,
the gapless approximation becomes invalid for $t_\times > N$, 
or for $\widetilde{\mu} < N^{-2/3}$. On the other side of the critical line
where $\widetilde{\mu}<0$ and $v_0 >0$,  $ \overline{h}(t)$ again equals
 $h_c(t)$ for $t \lesssim t_{\times }$, before changing to the
linear dependence for larger values of $t$. The statistics of global alignment
applies to the linear phase on scales $t>t_{\times }$. 

The above behavior of 
$ \bh(t) $ can be summarized by the homogeneous
scaling relation
\begin{equation}
 \bh(t;\widetilde{\mu}) 
= |\widetilde{\mu}|^{-1/2} f_\pm(t\,|\widetilde{\mu}|^{3/2}),
\label{scaling}
\end{equation}
with the two branches of the scaling functions $f_\pm(x)$ 
having the limiting forms $f_\pm(x\ll 1) =b\, x^{1/3}$,
$f_+(x\gg 1) \to {\rm const}$ for $\widetilde{\mu}>0$
and $f_-(x\gg 1) \propto x$ for $\widetilde{\mu}<0$~\footnote{
It should be noted that 
the scaling form (\ref{scaling}) 
also describes the score $ h_{\rm gapless}(t;\widetilde{\mu})$ of 
the zero-dimensional problem of gapless local alignment 
in the vicinity of its phase transition,
but with modified exponents. 
The results
 $ h _{\rm gapless}(t;\widetilde{\mu}) = |\widetilde{\mu}|^{-1}
g_\pm(t|\widetilde{\mu}|^2)$, with $g_\pm(x\ll 1) \propto x^{1/2}$, 
$g_+(x\gg 1) \to {\rm const}$ and $g_-(x\gg 1) \propto x$ can be
straightforwardly verified; see Ref.~\cite{mh} for a detailed discussion.}.
Such scaling relations
are widely encountered in physical systems undergoing continuous 
phase transitions  and have been studied
extensively by the modern theory of critical 
phenomena~\cite{phase-trans}.

\begin{figure}[t]
\centerline{\,\epsfig{file=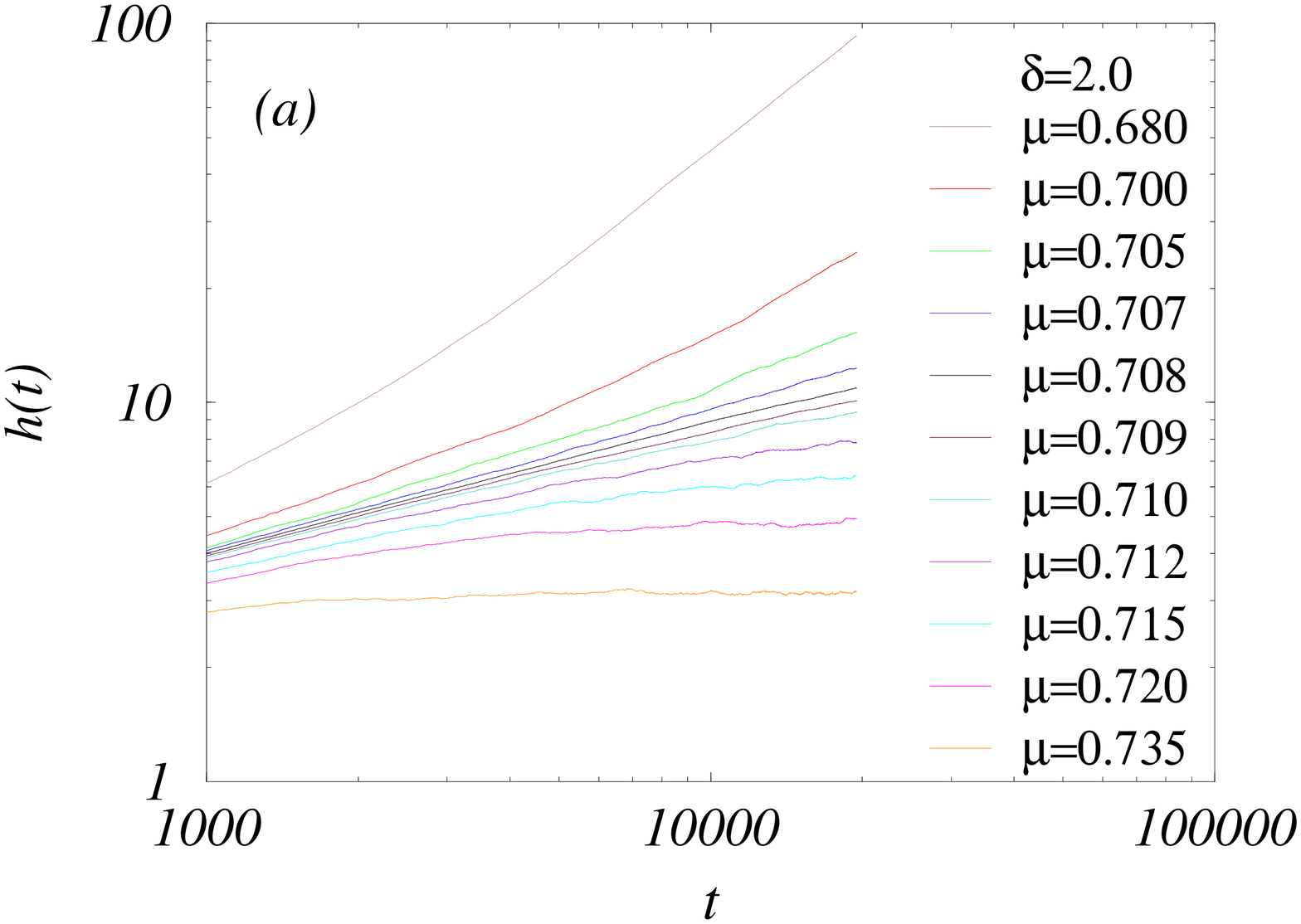, height=3.0in, angle=270}}
\centerline{\epsfig{file=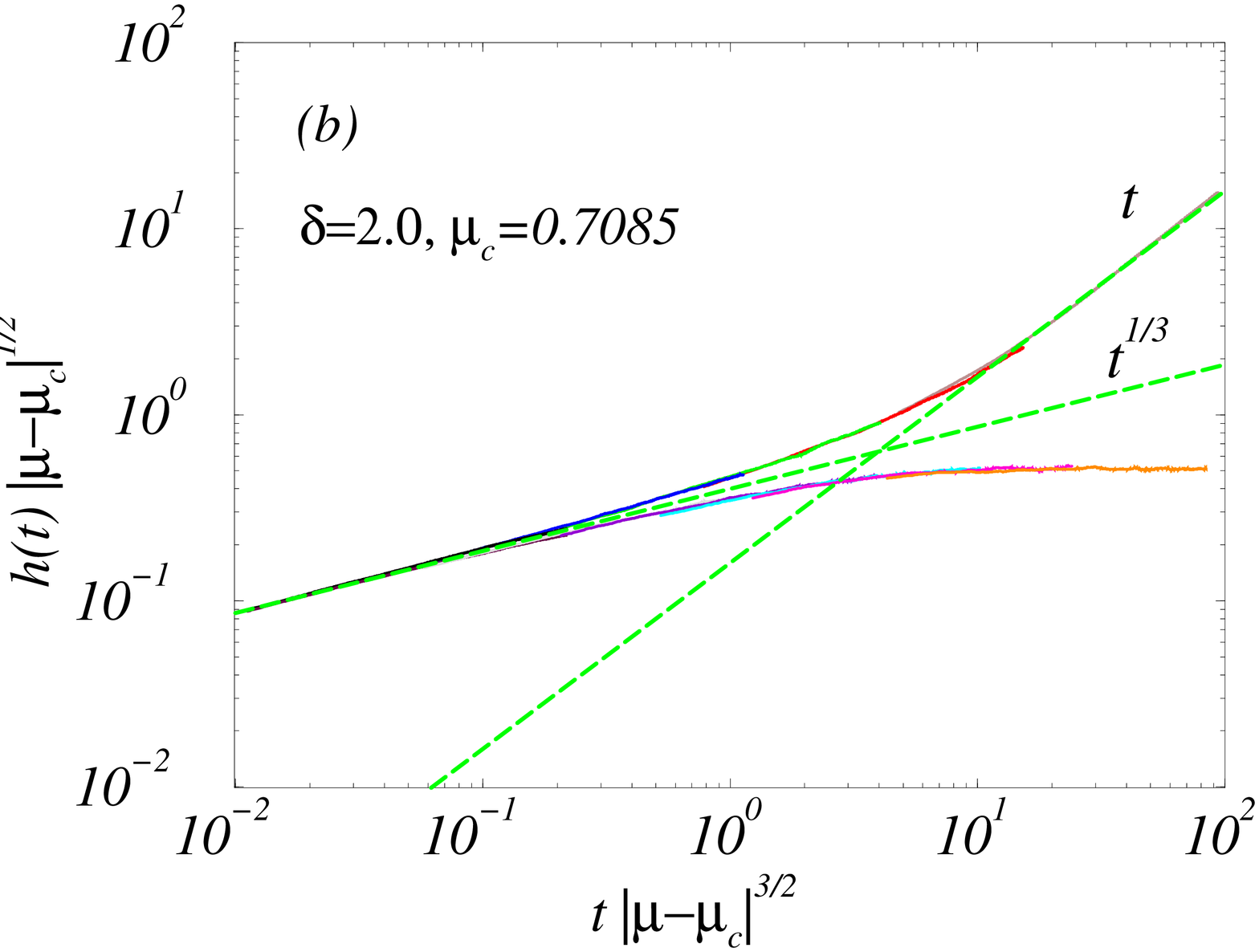, height=3.0in, angle=270}}
\vspace{\baselineskip}
\caption{
(a) The ensemble averaged score $ \bh(t) $ obtained
from the local alignment of random sequence pairs, with $\delta=2.0$,
and various values of $\mu$ slightly above and below the critical value
$\mu_c(\delta=2.0)\approx 0.7085$.
(The order of the curves corresponds to the order shown in the legend,
with the top curve having the  smallest value of $\mu$ and the bottom
curve having the largest value of $\mu$.)
Each curve is averaged over
$1000$ pairs of sequences of length $10,000$ each.
(b) The curves in (a) plotted according to the homogeneous scaling
form (see text). The dashed lines indicate the anticipated power law
forms in the two different regimes.
}
\end{figure}

In Fig.~6(a), we show the score average $\overline {h}(t)$
for local alignment of $1000$ 
random sequence pairs of length $10000$, for $\delta =2$
and various $\mu $'s taken $\pm 5\%$ around the critical value 
$\mu_c\approx 0.7085$.
 The anticipated scaling form (\ref{scaling}) 
can be checked by multiplying the horizontal axis of Fig.~6(a) 
by a factor $|\widetilde{\mu}|^{3/2}$ and the vertical axis by 
$|\widetilde{\mu}|^{1/2}$ individually for each curve. The result is
shown in Fig.~6(b). The $11$ curves displayed in Fig.~6(a) (each
containing $10^5$ data points) collapse   into two branches,
corresponding to the two branches of the scaling function $f_\pm$. 
The upper branch ($f_-$) describes the crossover of the 
critical behavior $h_c(t)$ to the linear behavior, and the lower branch 
($f_+$) describes the crossover of the critical behavior towards saturation. 
The only fitting parameter for this data collapse is the
location of the transition point $\mu _c$.

A similar scaling relation exists for the roughness of the score profile.
In Fig.~7(a), the width $\overline{w}(t)$ is plotted 
for various values of $\mu$, approaching the
phase transition from the log side. 
The data can be collapsed by  the same transformation as in
 Fig.~6(b), with the same fitting parameter $\mu_c$. 
The results (Fig.~7(b)) show clearly that 
$w(t;\widetilde{\mu})$ has the same form as $ \bh(t;\widetilde{\mu})$
in the log phase, i.e., the lower branch of Fig.~6(b). In particular, 
$w(t>t_\times)$ saturates to a constant value 
$w_{\rm sat}$  of the same order as $h_{\rm sat}$, yielding
\begin{equation}
w_{\rm sat}\sim b\, t_\times^{1/3}.\label{wsat}
\end{equation} 
This is just the expression
(\ref{width.scale}) describing the score roughness of {\em global} alignment, 
evaluated at $t=t_\times$. It is a manifestation of the general result 
that local alignment corresponds to global 
alignment applied to the selected subsequences.

\begin{figure}[t]
\centerline{\quad\epsfig{file=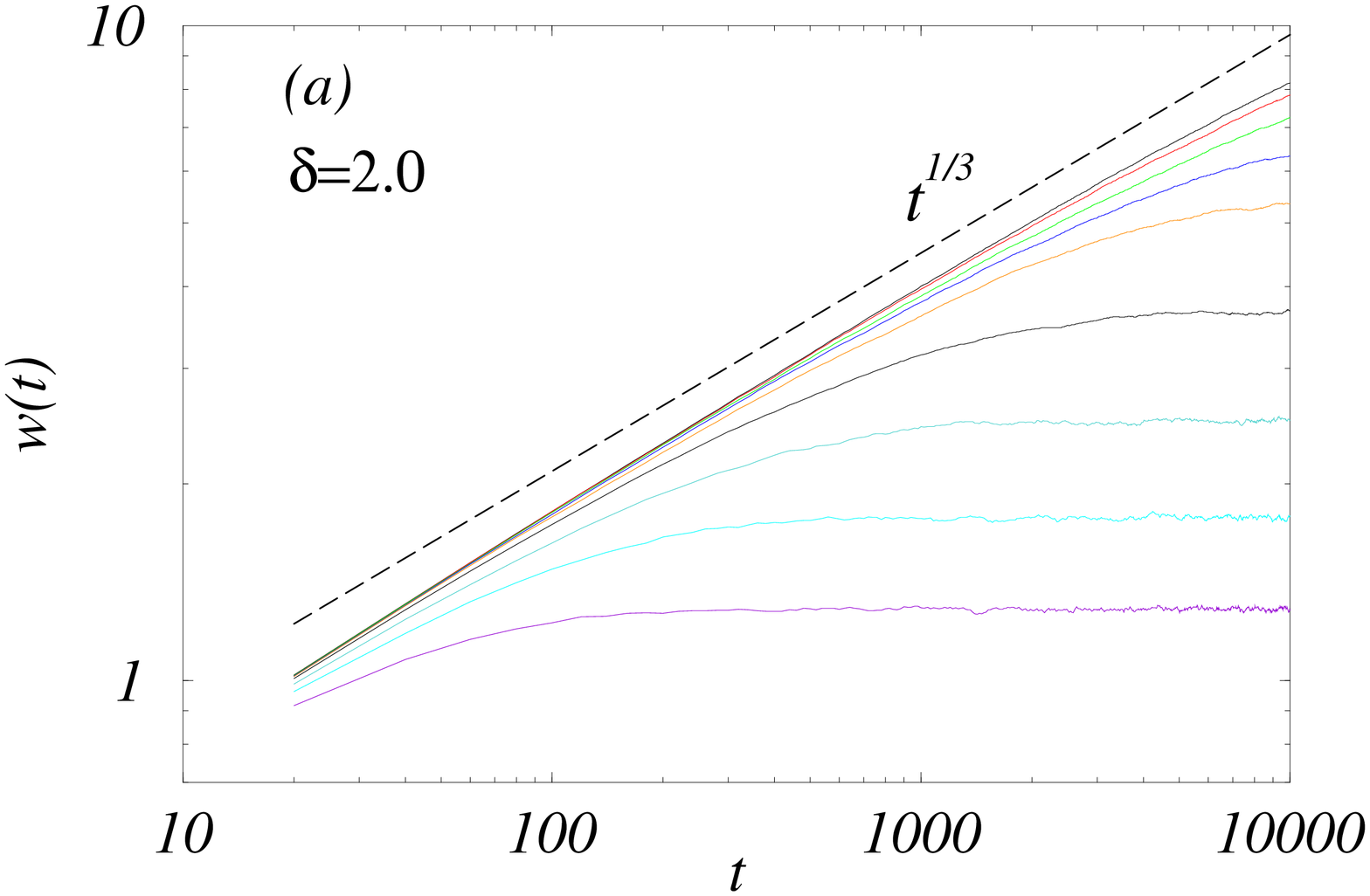, height=2.8in, angle=270}}
\centerline{\epsfig{file=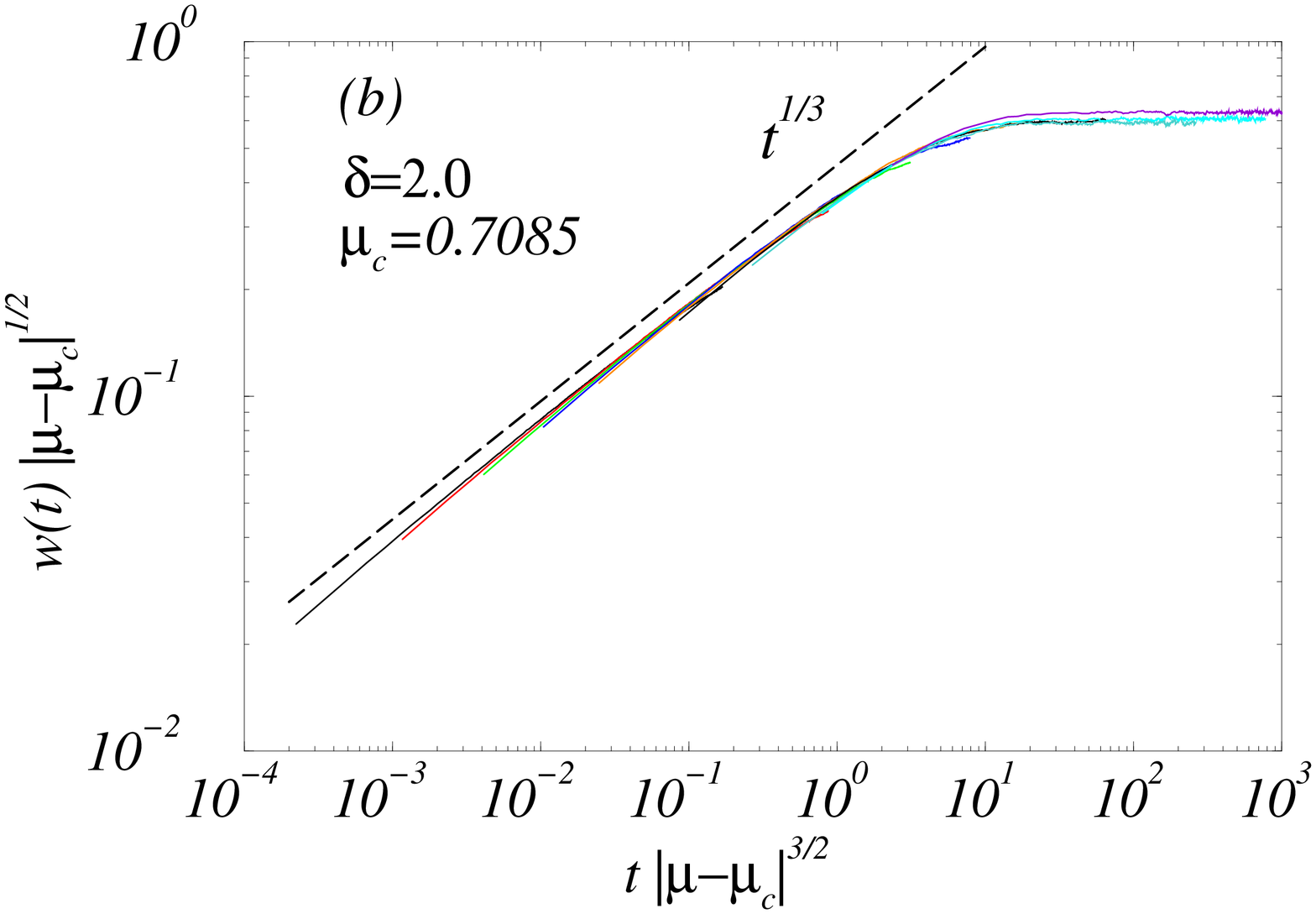, height=2.8in, angle=270}}
\caption{
(a) The ensemble averaged roughness $ \overline {w}(t) $ 
in the log phase, for $\delta=2.0$ and $\mu = 0.709$, $0.710$, $0.712$,
$0.715$, $0.720$, $0.735$, $0.765$, $0.825$, $0.950$ (from top to bottom). 
The sequences are the same as those used in Figs.~6.
(b) The roughness curves plotted according to the scaling form.
}
\end{figure}


\subsection{Correlated Subsequences: Similarity Detection and 
Parameter Dependence}

The existence of a phase transition in local alignment can be used to detect
sequence-sequence correlations:
Consider a pair of sequences ${\cal P}_1$ and ${\cal P}_2$ 
with mutually correlated subsequences located in the intervals  
from $i_0$ to $i_0+\ell/2 $ 
 and from $j_0$ to $j_0+\ell/2 $, respectively.
For concreteness, let the similarity of these subsequences
be generated by the evolution mechanism  described in Sec.~3.2,
with the strength of inter-sequence correlations characterized by
$\eps(\mu,\delta;p_s,p_t)$.  Our strategy 
for similarity detection is simply to choose 
 the scoring parameters $\delta$ and $\mu$ such
 that $v_0(\mu,\delta)<0$, i.e., the log phase is obtained if the sequences 
are uncorrelated, while keeping $v_0+\eps >0$,
so that the linear phase may be obtained instead for some duration of the 
correlated subsequences, $t_0<t<\ell$.
With this parameter choice, 
the profile $h(x,t)$ will have a constant background roughness $\wsat$
for  $t< t_0=i_0+j_0-1$, and a score peak signaling
subsequence correlations in the interval  $ t_0 < t < \ell$,
once 
\begin{equation}
(\epsilon + v_0) \cdot (t- t_0) > w_{\rm sat}. \label{cond3}
\end{equation}

Optimal similarity detection is obtained by maximizing the
 peak-to-background ratio,
\begin{equation}
\sigma = (\epsilon + v_0)/ \wsat \label{sig}
\end{equation}
which can be taken as a measure of the statistical significance of the 
 correlations detected.
Using the relations (\ref{wsat}) and (\ref{length}), we find
$\sigma \sim (\eps - |v_0|) \cdot |v_0|^{1/2}$, which is maximized at 
\begin{equation}
v^*_0 = v_0(\mu^*,\delta^*)=- \frac{\eps}{2}, \label{result1} 
\end{equation}
with
\begin{equation}
\sigma^* \propto \eps^{3/2} \sim t_c^{-1}. \label{result2}
\end{equation}

Eqs.~(\ref{result1}) and (\ref{result2}) are the central results of
this study. Eq. (\ref{result1}) shows that for weakly correlated
subsequences (i.e., $\eps \to 0$),  the optimal scoring parameters
should be close to the log side of the phase boundary. This is
a quantitative formulation of the empirical observation of
Vingron and Waterman~\cite{vw}. In addition to the choice of $v_0$,
Eq.~(\ref{result2}) shows that $t_c$ should still be minimized
as discussed in Sec.~3.2 for global alignment. 
These two conditions uniquely determine the optimal parameters 
$\mu^*$ and $\delta^*$.
Using the optimal result, the condition (\ref{cond3}) is reduced to 
$t-t_0 > t_c^*=t_c(\mu^*,\delta^*;p_s,p_t)$, which yields the
minimal subsequence length $\ell$ for which correlations can be
detected.


\section{SUMMARY}

In this study, we presented a statistical description of the 
Smith-Waterman local alignment algorithm,
focusing on the properties near the log-linear phase transition line.
We demonstrated how this knowledge can be exploited 
to provide quantitative criteria guiding the choices
of  alignment parameters for optimal detection of weak sequence correlations.
The optimal values $\mu^*$ and $\delta^*$ are obtained 
in terms of the substitution rate $p_s$ and the indel rate $p_t$
characterizing the statistics of sequence correlations. 
By analyzing the evolution of the spatially-extended score profile,
we are able to detect sequence correlations by a single run of the algorithm.
This is a very efficient way to optimize the scoring parameters 
empirically, and may be useful
in the alignment of a vast number of weakly
correlated sequences.


\section*{ACKNOWLEDGMENTS}

The authors have benefited from discussions with M.A. Mu\~{n}oz, D. Drasdo,
S.F. Altschul, and M.S. Waterman. T.H. acknowledges the financial support of a
research fellowship by the A.P. Sloan Foundation, and an young investigator
award from the Arnold and Mabel Beckman Foundation.





\end{document}